\documentclass[doublecol]{epl2} 

\begin{document}

\title{Quantum Hall Bilayer as Pseudospin Magnet}
\shorttitle{Quantum Hall Bilayer as Pseudospin Magnet} 

\author{O. Kyriienko\inst{1,2,3} \and K. Wierschem\inst{1} \and P. Sengupta\inst{1} \and I. A. Shelykh\inst{1,2,4}}
\shortauthor{O. Kyriienko \etal}

\institute{                    
  \inst{1} Division of Physics and Applied Physics, Nanyang Technological University 637371, Singapore\\
  \inst{2} Science Institute, University of Iceland, Dunhagi-3, IS-107, Reykjavik, Iceland\\
  \inst{3} Niels Bohr Institute, University of Copenhagen, Blegdamsvej 17,DK-2100 Copenhagen, Denmark\\
  \inst{4} ITMO University, St. Petersburg 197101, Russia}
  


\abstract{
We revisit the physics of electron gas bilayers in the quantum Hall regime [Nature, 432 (2004) 691; Science, 305 (2004) 950], where transport and tunneling measurements provided evidence of a superfluid phase being present in the system. Previously, this behavior was explained by the possible formation of a BEC of excitons in the half-filled electron bilayers, where empty states play the role of holes. We discuss the fundamental difficulties with this scenario, and propose an alternative approach based on a treatment of the system as a pseudospin magnet. We show that the experimentally observed tunneling peak can be linked to the XY ferromagnet (FM) to Ising antiferromagnet (AFM) phase transition of the $S=1/2$ XXZ pseudospin model, driven by the change in total electron density. This transition is accompanied by a qualitative change in the nature of the low energy spin wave dispersion from a gapless linear mode in the XY-FM phase to a gapped, quadratic mode in the Ising-AFM phase.}

\maketitle

\section{Introduction} The end of the 20th century was marked by several major achievements in the field of condensed matter physics, related to studies of unconventional states of matter. A significant breakthrough was the experimental observation \cite{Klitzing1980,Tsui1982,Stromer1983} and theoretical explanation of the quantum Hall effect (QHE). Although the integer quantum Hall effect can be explained using the concept of non-interacting fermions \cite{Laughlin1981,Kazarinov1982,Buttiker1988}, the explanation of the fractional quantum Hall effect is based on a non-perturbative treatment of the electron-electron interactions, resulting in the formation of a new state of matter---a strongly correlated incompressible quantum fluid \cite{Laughlin1983,Haldane1983,Jain1989}. Another discovery was an experimental realization of the Bose-Einstein condensation (BEC) of cold atoms~\cite{Anderson1995,Davis1995,Bradley1995,Andrews1995}. This latter achievement stimulated the search for BEC in solid state systems, where condensation of various bosonic quasiparticles, including magnons \cite{Democritov2006}, exciton-polaritons \cite{Kasprzak2006,SvenNature}, indirect excitons \cite{Butov2001,SnokeScience,High2012}, cavity photons \cite{Klaers2010}, and others was experimentally reported recently. In this vein, a remarkable setup consisting of electron bilayers in strong perpendicular magnetic fields was suggested as a system in which the physics of the QHE and BEC meet \cite{MacDonald2004,Eisenstein2004}.

The system under investigation consists of two quantum wells (QWs) with $n$-type conductivity placed in a strong perpendicular magnetic field $B$ which is tuned to make the total filling factor of the lowest Landau level (LL) equal to one, $\nu_T=1$. Interestingly, a bilayer system possesses a number of properties indicating the formation of a strongly correlated quantum state different from the states previously observed in monolayer QH systems. First, the tunneling between layers as a function of the interlayer voltage $V$ was shown to be qualitatively different for small and large electron concentrations \cite{Spielman2000}. Samples with high electron concentration demonstrated the well-known Coulomb suppression of tunneling at $V=0$ \cite{Eisenstein1992}. On the contrary, samples with low density exhibited a pronounced maximum at $V=0$, similar to that characteristic of the Josephson effect. Second, in counterflow experiments where the Hall voltages in individual QWs were measured separately as a function of the magnetic field, it was found that for values of the magnetic field corresponding to $\nu_T=1$, the Hall voltages in the counterflow experiment dropped to zero \cite{Kellog2004,Tutuc2004}.

The described experimental results were qualitatively explained in Ref. \cite{MacDonald2004} as consequences of BEC of excitons in electron bilayers. Indeed, in a bilayer system with total filling factor $\nu_T=1$ the electrons can be redistributed between the two layers in different ways; for example, one can imagine that all of them lie in the lower layer, and the upper layer is empty. This state was taken to be a vacuum state. Then, if one removes an electron from the lower layer and places it in the upper layer, one creates an excitation in the system, which is expected to behave as a boson. This situation is analogous to the one in the QHE ferromagnet \cite{Doretto}, with the difference being that the role of the real spin is played by a pseudospin describing the localization of the electron in the upper or lower layer. Then, to minimize the total energy of the electron-electron interactions, one needs to redistribute the electrons equally between the layers. From the point of view of the above defined vacuum state, this corresponds to the creation of $N_\phi$ excitons in the system, with $N_\phi=eBS/2\pi\hbar$ being the number of available states in a Landau level, where $e$ denotes electron charge and $S$ is the area of the sample. At low temperatures, these excitons undergo a transition to a condensed state characterized by the onset of superfluidity and the appearance of a gapless Bogoliubov mode in the spectrum of elementary excitations \cite{Cristiana}.
Further studies of QHE exciton concept were performed, showing its analogy to composite boson or 111 Quantum Hall state \cite{Simon2003,Moller2008,Milovanovic2009,Papic2012}. In particular, along these lines the transition between composite boson and composite fermion states of two decoupled Fermi liquids were studied \cite{Simon2003}.

It should be noted, however, that the approach introduced in Ref. \cite{MacDonald2004} faces a number of fundamental difficulties. The first one is connected to the choice of the vacuum state. This does not generate any controversy for the quantum Hall ferromagnet (QHF), for which the vacuum corresponds to the state in which all spins are aligned along the magnetic field and the Zeeman energy is minimal. However, for the bilayer system, if one accounts for the possibility (however small) of tunneling between the two wells, the state having the minimal energy will evidently corresponds to the case where the wavefunction of the electron is a symmetric combination of the wavefunctions localized in the upper and lower wells for which the electrons are equally redistributed between the wells. Thus the state with all electrons concentrated in one of the wells can by no means be taken to be the vacuum and the concept of condensing bosons becomes shaky.

The second difficulty concerns the treatment of the excitations in the QHF as non-interacting or weakly interacting bosons which is possible only when the number of excitations $N$ in the system is much less then the total number of states in a Landau level, $N\ll N_\phi$. This condition is clearly violated in the case of a quantum Hall bilayer (QHB) when $N=N_\phi/2$.

In the present Letter we use an alternative phenomenological model for the description of the quantum Hall bilayer at total filling factor $\nu_T=1$. We show that the experimentally observed superfluid behavior of the QHB system can be explained within the pseudospin model \cite{Moon1995} and is associated with the XY-FM ground state phase characterized by non-zero spin stiffness and gapless linear dispersion of elementary excitations.

\section{Pseudospin description} 
\begin{figure}
\centering
\includegraphics[width=0.9\linewidth]{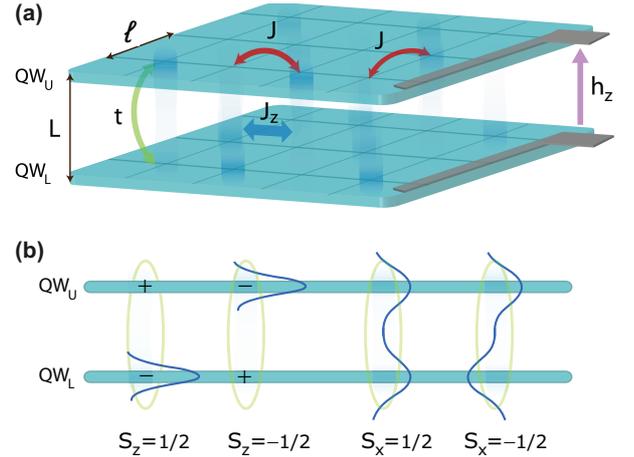}
\caption{(color online) (a) Sketch of the system, showing two quantum wells (QW$_U$ and QW$_L$) separated by distance $L$. The layers are in the quantum Hall effect regime, where effective lattice spacing is given by magnetic length $\ell$. $J_z$, $J$,  $t$ and $h_z$ correspond to the effective couplings and magnetic fields in the pseudospin description. The Ising term $J_z$ describes the difference in energy between parallel and antiparallel electric dipoles; the term $J$ describes spin exchange processes of the exchange of electrons between two neighboring orbitals; the term $h_z$ describes the effect of the external gate voltage creating an asymmetry between the upper and lower QWs; and the term $t$ describes tunneling between layers. (b) A schematic illustration of the states having pseudospins $S_z=\pm1/2$ and $S_x=\pm1/2$.}
\label{fig:sketch}
\end{figure}

The system under consideration consists of two thin quantum wells (QWs), which contain a two-dimensional electron gas in a strong perpendicular magnetic field [see sketch in Fig. \ref{fig:sketch}(a)]. We consider the situation where in-plane magnetic field is absent and thus real spin-related effects can be neglected \cite{Giudici2008,Giudici2010,Finck2010}. For non-interacting particles, the eigenstates of the system corresponding to the first Landau level are a set of circular orbitals with radius $\ell = \sqrt{\hbar/eB}$, whose guiding centers form a regular grid. In the present work, we concentrate on the case of total filling factor $\nu_T=1$, where every orbital is occupied by a single electron representing a two-level system, which can be mapped to a $S=1/2$ pseudospin. We denote the states with an electron being localized in the upper and lower wells as having $S_z=+1/2$ and $S_z=-1/2$ [Fig. \ref{fig:sketch}(b)]. The symmetric and antisymmetric states correspond to an orientation of the pseudospin along the $x$ axis, $S_x=\pm1/2$.

To construct the model Hamiltonian of the system in pseudospin representation let us note the following. The direct and exchange Coulomb interactions between electrons lead to the appearance of effective interactions between pseudospins, which can be of Ising ($J_z$) and spin exchange ($J$) type, whose meaning is clarified in Fig.~\ref{fig:sketch}. The Ising term corresponds to dipole-dipole interaction and is of an antiferromagnetic nature, $J_z > 0$, and the exchange interaction is of ferromagnetic type, $J < 0$. Tunneling between layers leads to a hybridization of the two modes, acting as an effective transverse magnetic field $t$ lying in the $x$-$y$ plane. Finally, the applied voltage between layers creates an asymmetry between the upper and lower QWs, thus leading to the emergence of an effective longitudinal field $h_z$ along the $z$ axis.

The system can thus be described by the two-dimensional $S=1/2$ XXZ model Hamiltonian:
\begin{equation}
\hat{\mathcal{H}} = J\sum_{\langle i,j \rangle} \left[ \widehat{S}_i^x \widehat{S}_j^x + \widehat{S}_i^y \widehat{S}_j^y + \Delta \widehat{S}_i^z \widehat{S}_j^z \right] - h_z \sum_{i}\widehat{S}_i^z - t\sum_{i}\widehat{S}_i^x,
\label{eq:H}
\end{equation}
where $\Delta = J_z/J$ denotes the anisotropy parameter that depends on the experimental configuration (see discussion below). The operators $\widehat{S}_{i,j}^{x,y,z}$ are standard pseudospin $1/2$ operators with commutation relations defined as $[\widehat{S}_{i}^{\alpha},\widehat{S}_{j}^{\beta}]=\epsilon_{\alpha\beta\gamma}\delta_{ij}\widehat{S}_{i}^{\gamma}$. We  neglect the states containing empty orbitals or orbitals carrying two electrons (one in the upper well and one in the lower well) and drop the spin-independent term in the interaction energy which does not influence the results. We note that the microscopic derivation of the generic pseudospin model was performed in Ref. \cite{Burkov2002}. However, in the following we do not consider the full SU(4) spin representation, where both real spin and pseudospin degree of freedom are accounted for, and concentrate only on the latter.

First let us consider uncoupled layers for which the effective transverse field vanishes, $t/J \rightarrow 0$, and study the longitudinal field dependence of the pseudospin system. We are primarily interested in the ground state properties of the quantum system, and want to answer the question: is the ground state gapless or gapped? While the latter will manifest itself in a finite resistivity of the drag measurement in counter-flow experiments, the former can account for the superfluid-like behavior of the system observed in Ref. \cite{Spielman2000}. To answer the question, several strategies can be used. First, one can use a quantum Monte Carlo approach which was shown to be a universal tool for studying spin systems, in particular for extracting phase boundaries and corresponding critical exponents of a model. At the same time, the estimation of ground state behavior can also be done using a spin wave theory, while being less quantitatively successful for calculation of other observables. In the following, we have used the first method.

We have simulated the Hamiltonian (\ref{eq:H}) on square lattices using the Stochastic Series Expansion (SSE) quantum Monte Carlo (QMC) method with standard operator loop updates for the longitudinal field \cite{Sandvik1999,Syljuasen2002}. The simulations were performed on finite size lattices of the form $L \times L$ ($12 \leq L \leq 24$). Estimates for ground state properties were obtained from a finite-size and finite-temperature scaling analysis of the obtained data. To characterize the different phases, we compute the uniform and staggered magnetisation as well as the spin stiffness which provides a convenient way to detect the presence or absence of a spin gap in the ground state.
\begin{figure}
\centering
\includegraphics[width=0.7\linewidth]{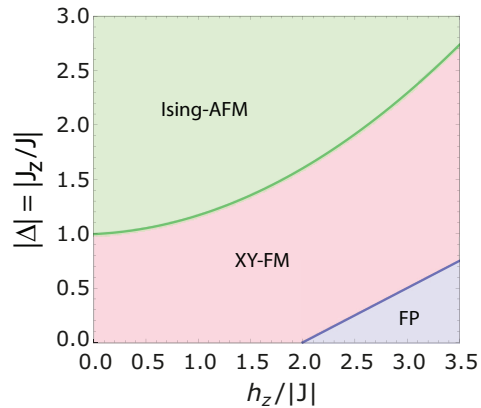}
\caption{(color online) Phase diagram of the pseudospin system plotted as a function of absolute value of the dimensionless anisotropy parameter $\Delta$ and dimensionless longitudinal magnetic field $h_z/J$. We see that for small anisotropy ($|\Delta| < 1$) and longitudinal field $h_z/J < 2$ the ground state state is gapless ($\rho_s \neq 0$), and represents a superfluid phase.}
\label{fig:hz}
\end{figure}

The ground state phase diagram of the system is presented in Fig. \ref{fig:hz}. We encounter three phases, two of which are gapped (Ising-AFM and fully polarized phases), and one is gapless (XY-FM phase). From the point of view of the experimental configuration relevant to the onset of superfluidity in counterflow experiments, the most important feature here is the Ising to XY-FM phase transition. For $h_z > 0$, this phase can be understood in terms of a BEC of field induced magnetic excitations (magnons), and is different from the proposed excitonic BEC scenario. The transition to the fully polarized (FP) phase occurs only at high magnetic fields $h_z/|J| >2$, and the corresponding boundary is given by the analytical expression $(\Delta/J)^{cr}=h_z/2 - 1$ \cite{Yunoki2002} and according to our estimations corresponds to a parameter range which is not covered in current experiments. 

A large transverse effective magnetic field, characterized by the parameter $t$, will lead to the appearance of a gapped ground state of the system, and thus destroy any superfluid order. However, we note that the tunneling matrix element between upper and lower wells $t$ rapidly decreases with the distance between them, $L$, and the value of the separating potential barrier, $U_0$. It can be estimated as $t\approx 4U_0e^{-\sqrt{2mU_0}L/\hbar}$, and is typically orders of magnitude smaller as compared to the interaction parameters $J_z$ and $J$ in experiment \cite{Spielman2000}.
\begin{figure}
\centering
\includegraphics[width=1.0\linewidth]{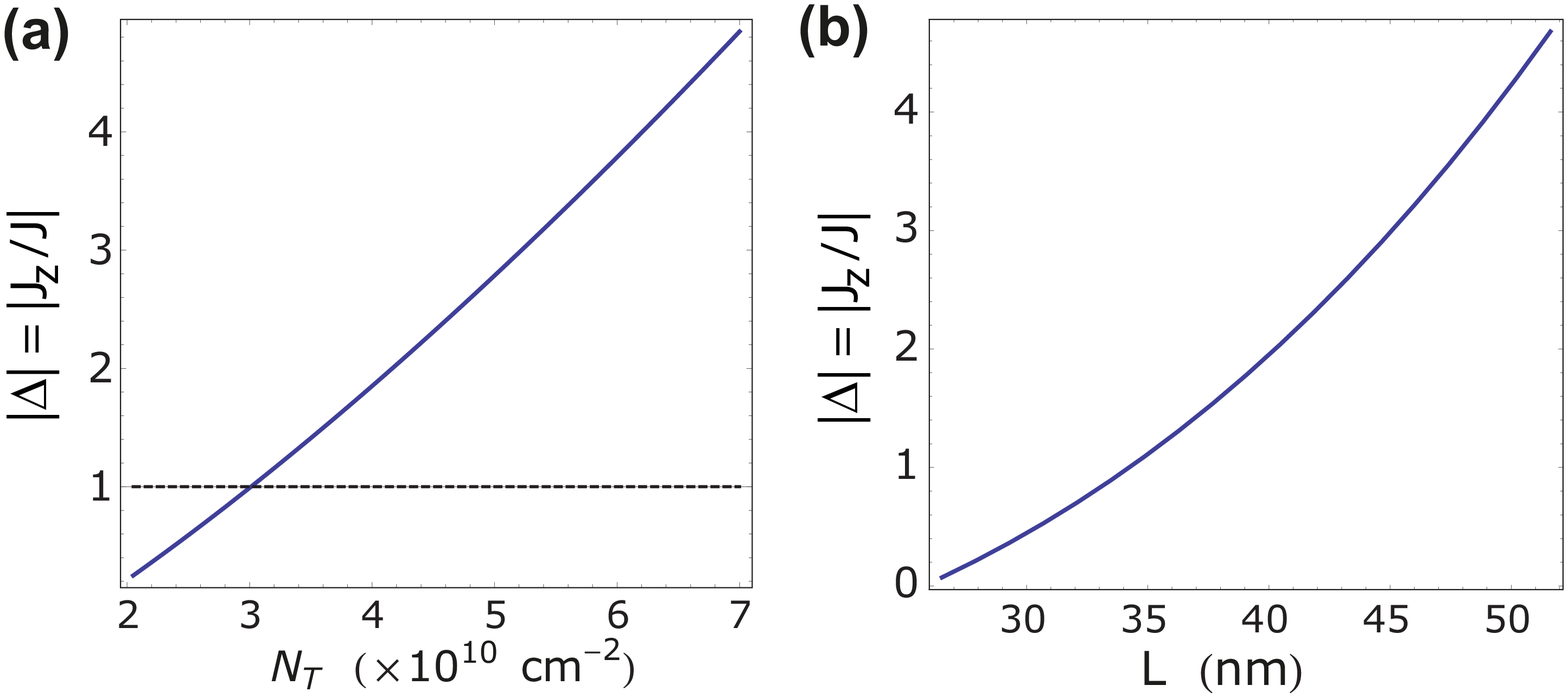}
\caption{(color online) Interaction anisotropy constant $|\Delta|$ as a function of total electron density (a) and layer separation $L$ (b). In plot (a) the distance between QW centers is fixed to $L = 27.9$ nm. The electron density in (b) is fixed to $N_T = 2\times 10^{10}$ cm$^{-2}$.}
\label{fig:Delta}
\end{figure}

\section{Relation to experiment} To relate the phase diagram obtained above to the experimental configuration, let us recall some details about the quantum Hall bilayer system, which was studied experimentally \cite{Spielman2000}. It consists of two GaAs quantum wells of width $L_{QW}=18$ nm, separated by an Al$_{0.9}$Ga$_{0.1}$As barrier with thickness $L_{spacer}=9.9$ nm. This leads to a separation of $L= 27.9$ nm between the centers of the QWs. The total 2D concentration is controlled by gate electrodes and varies from $N_T^{(0)}= N_{1}^{(0)} + N_{2}^{(0)} = 10^{11}$ cm$^{-2}$ to $N_T^{(0)}= 10^{10}$ cm$^{-2}$. Additionally, an external voltage applied to the layers can make electron concentration in the QWs unequal, $N_1 \neq N_2$, thus introducing a longitudinal effective magnetic field $h_z$ in the pseudospin description.

The relevant parameter of the system is a magnetic length $\ell = \sqrt{\hbar/eB}$, which controls the spacing between effective lattice sites and, consequently, the interaction between electrons. Typically in QHB experiments the total filling factor $\nu_T = 2\pi \hbar N_T / eB$ is fixed to unity by a fine tuning of the magnetic field, leading to an unambiguous relation between magnetic length and electron density, $\ell = \sqrt{1/2\pi N_T}$. The transition from dissipative to superfluid transport was experimentally reported when the electron density was decreased.

To develop a quantitative description of QHB within the pseudospin model and compare the results with those obtained in experiment, we calculate the constants $J_z$ and $J$ as matrix elements of the Coulomb interaction for the states of the two electrons located at the orbitals with neighbouring guiding centers. The procedure is described in the Supplemental Material \cite{SM}, and allows us to calculate the anisotropy constant $\Delta = J_z/J$ as a function of the total electron density [Fig. \ref{fig:Delta}(a)] and interlayer separation $L$ [Fig. \ref{fig:Delta}(b)]. In the plots we show the absolute value of the anisotropy constant, as the calculated exchange constant $J$ has negative sign. This results in a ferromagnetic nature of XY phase. Notably, the behavior of ferromagnetic phase ($J<0$) of XY limit ($|\Delta|<1$) is not qualitatively different from that of the antiferromagnetic case, $J>0$, both corresponding to gapless superfluid behavior.

As can be seen from Fig. \ref{fig:Delta}(a), in the absence of tunneling between the layers and layer asymmetry ($h_z=t=0$), the XY-FM to Ising phase transition occurs at the isotropic Heisenberg point $|\Delta| = 1$ which corresponds to a density $N_T = 3. \times 10^{10}$ cm$^{-2}$. In experiment, superfluid behavior was reported for a density $N_T^{cr} = 5.4 \times 10^{10}$ cm$^{-2}$ \cite{Spielman2000}, which is quite close to our result. The difference can be attributed to the approximative nature of our model, where we considered thin QWs and neglected the long-range nature of Coulomb interaction between electrons at different orbitals. This for instance can lead to frustration and emergence of a spin liquid phase \cite{Burkov2002}. We estimated the value of a next nearest neighbour Ising interaction $J_z'$ to be seven times smaller than $J_z$ for densities corresponding to the transition point, and note that this also corresponds to a superfluid phase in the extended Heisenberg model \cite{Hebert2001}.

\section{Dispersion of excitations} In the relevant limit of small effective magnetic fields $h_z$ and $t$, it is possible to derive the dispersions of elementary excitations in the quantum Hall bilayer system. They correspond to spin wave excitations, or magnons.

We use the standard spin wave analysis based on Holstein-Primakoff transformation \cite{Holstein1940,SandvikNotes,Joannopoulos1987,Hamer1991} to study the low energy magnon excitations of the Hamiltonian (\ref{eq:H}) with $t=h_z=0$ (see \cite{SM} for details). The resulting dispersion is
\begin{equation}
\label{eq:omega}
\omega(\mathrm{k}) = 2 |J| \left\{\begin{array}{ll}
     \sqrt{(1-\gamma_{\mathrm{k}}) ( 1 + |\Delta| \gamma_{\mathrm{k}})}, \hbox{$|\Delta| \leq 1$,} \\
     \sqrt{\Delta^2 - \gamma_{\mathrm{k}}^2}, \hbox{$\Delta \geq 1$,}
   \end{array}
\right.
\end{equation}
where $\gamma_{\mathrm{k}} = [\cos k_x + \cos k_y]/2$ and momenta $k_x$, $k_y$ are measured in units of the inverse lattice constant, $2\pi/\ell$.
\begin{figure}
\centering
\includegraphics[width=0.9\linewidth]{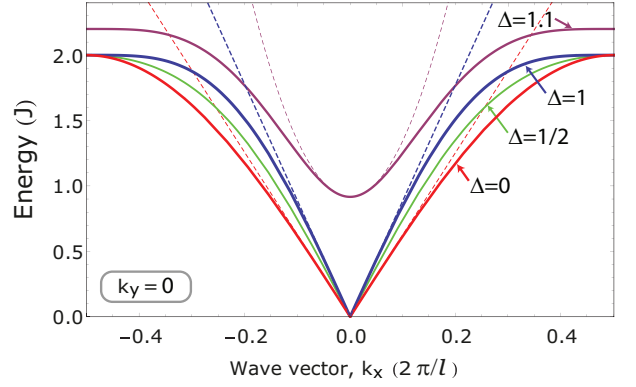}
\caption{(color online) Dispersions of the QHB magnons as a function of wave vector $k_x$, plotted for the $k_y = 0$ wave vector value.}
\label{fig:disp}
\end{figure}

Using equation (\ref{eq:omega}) we can plot the dispersions of the spin wave excitations in the full range of the anisotropy parameter $\Delta$. The results are depicted in Fig. \ref{fig:disp}, where we show the spin wave energy as a function of wave vector along the $x$ axis, $k_x$, while fixing the wave vector $k_y = 0$. The magnons are found to be gapless for $\Delta = 0,~1/2,~1$.
They correspond to the gapless Goldstone modes in the superfluid phase of the weakly interacting BEC. The experimental observation of gapless excitations can thus be attributed to spontaneous symmetry breaking in the XY model, leading to superfluid pseudospin behavior. The magnon velocity $v$, defined as $\omega(\mathrm{k}) = v k$, can be calculated from the expansion around $(k_x, k_y) = (0, 0)$ point, and is equal to $v = |J| \sqrt{1+|\Delta|}$. The resulting approximate relations are shown by dashed lines in Fig. \ref{fig:disp}. As seen in a similar expansion for the Ising limit $|\Delta| > 1$, the spin wave dispersion becomes gapped with $\omega(\mathbf{k}) = 2 J \sqrt{\Delta^2 -1} + J/(2 \sqrt{\Delta^2 -1}) k^2$.

\section{Conclusions} To summarize, we revisit the system of the electron quantum Hall bilayer, where superfluid behavior of the system has been experimentally observed. We show that the existing description of the system in terms of excitons faces fundamental problems. As an alternative, we propose a quantitative theory based on a model 2D pseudospin magnet, that explains the observed behavior as a consequence of the onset of a superfluid phase in the XY limit of the Heisenberg model. The calculated critical electron density at which the gapped to gapless transition occurs is consistent with experimental measurements. We analyze the dispersions of the elementary excitations in the system, which correspond to spin waves (magnons), and demonstrate that for relevant values of the parameters it corresponds to a Goldstone mode satisfying the Landau criterion of superfluidity \cite{LegettBook}.

\acknowledgments
We thank Dr. Inti Sodemann for valuable discussions. The work was supported by the Ministry of Education, Singapore under grants Tier 1 project ``Polaritons for novel device applications'' (O.K. and I.S.) and MOE2011-T2-1-108 (K.W. and P.S.) and FP7 IRSES project QOCaN, and Rannis project ``Bose and Fermi systems for spintronics''. O.K. acknowledges the support from Eimskip Fund.


\begin{thebibliography}{0}

\bibitem{Klitzing1980}
  \Name{von Klitzing K., Dorda G. \and Pepper M.}
  \REVIEW{Phys. Rev. Lett.}{45}{1980}{494}.
  
\bibitem{Tsui1982}
  \Name{Tsui D. C., Stormer H. L., \and Gossard A. C.}
  \REVIEW{Phys. Rev. Lett.}{48}{1982}{1559}.

\bibitem{Stromer1983}
  \Name{Stormer H. L., Chang A., Tsui D. C., Hwang J. C. M., Gossard A. C., \and Wiegmann W.}
  \REVIEW{Phys. Rev. Lett.}{50}{1983}{1953}.

\bibitem{Laughlin1981}
  \Name{Laughlin R. B.}
  \REVIEW{Phys. Rev.}{23}{1981}{5632}.

\bibitem{Kazarinov1982}
  \Name{Kazarinov R. F. \and Luryi S.}
  \REVIEW{Phys. Rev. B}{25}{1982}{7626}.

\bibitem{Buttiker1988}
  \Name{Buttiker M.}
  \REVIEW{Phys. Rev. B}{38}{1988}{9375}.

\bibitem{Laughlin1983}
  \Name{Laughlin R. B.}
  \REVIEW{Phys. Rev. Lett.}{50}{1983}{1395}.

\bibitem{Haldane1983}
  \Name{Haldane F. D. M.}
  \REVIEW{Phys. Rev. Lett.}{51}{1983}{605}.

\bibitem{Jain1989}
  \Name{Jain J. K.}
  \REVIEW{Phys. Rev. Lett.}{63}{1989}{199}.

\bibitem{Anderson1995}
  \Name{Anderson M. H., Ensher J. R., Matthews M. R., Wieman C. E. \and Cornell E. A.}
  \REVIEW{Science}{269}{1995}{198}.

\bibitem{Davis1995}
  \Name{Davis K. B., Mewes M.-O., Andrews M. R., van Druten N. J., Durfee D. S., Kurn D. M.,  \and Ketterle W.}
  \REVIEW{Phys. Rev. Lett.}{75}{1995}{3969}.

\bibitem{Bradley1995}
  \Name{Bradley C. C., Sackett C. A., Tollett J. J., \and Hulet R. G.}
  \REVIEW{Phys. Rev. Lett.}{75}{1995}{1687}.

\bibitem{Andrews1995}
  \Name{Andrews M. R., Townsend C. G., Miesner H.-J., Durfee D. S., Kurn D. M. \and Ketterle W.}
  \REVIEW{Science}{275}{1995}{637}.

\bibitem{Democritov2006}
  \Name{Demokritov S. O., Demidov V. E., Dzyapko O., Melkov G. A., Serga A. A., Hillebrands B. \and Slavin A. N.}
  \REVIEW{Nature (London)}{443}{2006}{430}.

\bibitem{Kasprzak2006}
  \Name{Kasprzak J. \textit{et al.}}
  \REVIEW{Nature (London)}{443}{2006}{409}.

\bibitem{SvenNature}
  \Name{Schneider C. \textit{et al.}}
  \REVIEW{Nature (London)}{497}{2013}{348}.

\bibitem{Butov2001}
  \Name{Butov L. V., Ivanov A. L., Imamoglu A., Littlewood P. B., Shashkin A. A., Dolgopolov V. T., Campman K. L. \and Gossard A. C.}
  \REVIEW{Phys. Rev. Lett.}{86}{2001}{5608}.
  
\bibitem{SnokeScience}
  \Name{Snoke D.}
  \REVIEW{Science}{298}{2002}{1368}.  

\bibitem{High2012}
  \Name{High A. A., Leonard J. R., Hammack A. T., Fogler M. M., Butov L. V., Kavokin A. V., Campman K. L. \and Gossard A. C.}
  \REVIEW{Nature (London)}{483}{2012}{584}.

\bibitem{Klaers2010}
  \Name{Klaers J., Schmitt J., Vewinger F. \and Weitz M.}
  \REVIEW{Nature (London)}{468}{2010}{545}.

\bibitem{MacDonald2004}
  \Name{MacDonald A. \and Eisenstein J.}
  \REVIEW{Nature (London)}{432}{2004}{691}.

\bibitem{Eisenstein2004}
  \Name{Eisenstein J.}
  \REVIEW{Science}{305}{2004}{950}.

\bibitem{Spielman2000}
  \Name{Spielman I. B., Eisenstein J. P., Pfeiffer L. N. \and West K. W.}
  \REVIEW{Phys. Rev. Lett.}{84}{2000}{5808}.

\bibitem{Eisenstein1992}
  \Name{Eisenstein J. P., Pfeiffer L. N. \and West K. W.}
  \REVIEW{Phys. Rev. Lett.}{74}{1995}{1419}.

\bibitem{Kellog2004}
  \Name{Kellogg M., Eisenstein J. P., Pfeiffer L. N. \and West K. W.}
  \REVIEW{Phys. Rev. Lett.}{93}{2004}{036801}.

\bibitem{Tutuc2004}
  \Name{Tutuc E., Shayegan M. \and Huse D. A.}
  \REVIEW{Phys. Rev. Lett.}{93}{2004}{036802}.

\bibitem{Doretto}
  \Name{Doretto R. L., Caldeira A. O. \and Girvin S. M.}
  \REVIEW{Phys. Rev. B}{71}{2005}{045339}.

\bibitem{Cristiana}
  \Name{Doretto R. L., Caldeira A. O. \and Smith C. M.}
  \REVIEW{Phys. Rev. Lett.}{97}{2006}{186401}.

\bibitem{Simon2003}
  \Name{Simon S. H., Rezayi E. H. \and Milovanovic M. V.}
  \REVIEW{Phys. Rev. Lett.}{91}{2003}{046803}.

\bibitem{Moller2008}
  \Name{M\"{o}ller G., Simon S. H. \and Rezayi E. H.}
  \REVIEW{Phys. Rev. Lett.}{101}{2008}{176803}.

\bibitem{Milovanovic2009}
  \Name{Milovanovi\'{c} M. V. \and Papi\'{c} Z.}
  \REVIEW{Phys. Rev. B}{79}{2009}{115319}.

\bibitem{Papic2012}
  \Name{Papi\'{c} Z. \and Milovanovi\'{c} M. V.}
  \REVIEW{Mod. Phys. Lett. B}{26}{2012}{1250134}.

\bibitem{Moon1995}
  \Name{Moon K., Mori H., Yang K., Girvin S. M., MacDonald A. H., Zheng L., Yoshioka D. \and Zhang S.-C.}
  \REVIEW{Phys. Rev. B}{51}{1995}{5138}.

\bibitem{Giudici2008}
  \Name{Giudici P., Muraki K., Kumada N., Hirayama Y. \and Fujisawa T.}
  \REVIEW{Phys. Rev. Lett.}{100}{2008}{106803}.

\bibitem{Giudici2010}
  \Name{Giudici P., Muraki K., Kumada N. \and Fujisawa T.}
  \REVIEW{Phys. Rev. Lett.}{104}{2010}{056802}.

\bibitem{Finck2010}
  \Name{Finck A. D. K., Eisenstein J. P., Pfeiffer L. N. \and West K. W.}
  \REVIEW{Phys. Rev. Lett.}{104}{2010}{016801}.

\bibitem{Burkov2002}
  \Name{Burkov A. A. \and MacDonald A. H.}
  \REVIEW{Phys. Rev. B}{66}{2002}{115320}.

\bibitem{Sandvik1999}
  \Name{Sandvik A. W.}
  \REVIEW{Phys. Rev. B}{59}{1999}{R14157(R)}.

\bibitem{Syljuasen2002}
  \Name{Syljuasen O. F. \and Sandvik A. W.}
  \REVIEW{Phys. Rev. E}{66}{2002}{046701}.

\bibitem{Hebert2001}
  \Name{H\'{e}bert F., Batrouni G. G., Scalettar R. T., Schmid G., Troyer M. \and Dorneich A.}
  \REVIEW{Phys. Rev. B}{65}{2001}{014513}.

\bibitem{SM}
  See Supplemental Material at \textit{http://shelykhgroup.com/images/stories/smbilayers.pdf} for details of derivation of interaction constants and spin wave dispersions.


\bibitem{Yunoki2002}
  \Name{Yunoki S.}
  \REVIEW{Phys. Rev. B}{65}{2002}{092402}.

\bibitem{Holstein1940}
  \Name{Holstein T. \and Primakoff H.}
  \REVIEW{Phys. Rev.}{58}{1940}{1098}.

\bibitem{SandvikNotes}
  \Name{Sandvik A. W.}
  \REVIEW{AIP Conf. Proc.}{1297}{2010}{135}; doi: 10.1063/1.3518900.

\bibitem{Joannopoulos1987}
  \Name{Gomez-Santos G. \and Joannopoulos J. D.}
  \REVIEW{Phys. Rev. B}{36}{1987}{8707}.

\bibitem{Hamer1991}
  \Name{Hamer C. J., Oitmaa J. \and Weihong Z.}
  \REVIEW{Phys. Rev. B}{43}{1991}{10789}.

\bibitem{LegettBook}
  \Name{Legett A.}
  \Book{Condensation and Cooper Pairing in Condensed-Matter Systems}
  \Publ{Oxford University Press, Oxford}
  \Year{2006}.


\end{thebibliography}
\end{document}